\documentclass[reprint,pdflatex,amsmath,amssymb,aps,prd,floatfix]{revtex4-2}
\usepackage[english]{babel}
\usepackage{graphicx}
\usepackage{dcolumn}
\usepackage{bm}
\usepackage{appendix}
\usepackage{natbib}

\begin{document}

\title{Redshift space distortions in Lagrangian space and the linear large scale velocity field of dark matter}

\author{Emily Tyhurst$^1$}
 \email{etyhurst@physics.utoronto.ca}
\author{Hamsa Padmanabhan$^{2,1,7}$}
 \email{hamsa.padmanabhan@unige.ch}
\author{Ue-Li Pen$^{1,3,4,5,6, 7}$}

\affiliation{
 1. Canadian Institute for Theoretical Astrophysics.
  University of Toronto, Toronto, Ontario, Canada.}\affiliation{2. Université de Genève, 24 quai Ernest-Ansermet, CH 1211 Genève 4, Switzerland.}
  \affiliation{3. Academia Sinica Institute of Astronomy and Astrophysics in Taiwan, ASIAA, 11F of AS/NTU Astronomy-Mathematics Building, No.1, Sec. 4, Roosevelt Rd, Taipei 10617, Taiwan, R.O.C.}
\affiliation{4. Canadian Institute for Advanced Research. CIFAR Program in Gravitation and Cosmology. 661 University Ave, Toronto, Ontario, M5G 1Z8, Canada}\affiliation{5. Perimeter Institute for Theoretical Physics. 31 Caroline Street North, Waterloo, Ontario, N2L 2Y5, Canada}\affiliation{6. Max-Planck-Institut f\"ur Radioastronomie, Auf dem H\"ugel 69 53121, Bonn, Germany}\affiliation{7. Dunlap Institute for Astronomy and Astrophysics. University of Toronto, 50 St. George Street, Toronto, Ontario M5S 3H4, Canada}

\date{\today}

\begin{abstract}
Untangling the connection between redshift space coordinates, a velocity measurement, and three dimensional real space coordinates, is a cosmological problem that is often modeled through a linear understanding of the velocity-position coupling. This linear information is better preserved in the Lagrangian space picture of the matter density field. Through Lagrangian space measurements, we can extract more information and make more accurate estimates of the linear growth rate of the universe. In this paper, we address the linear modelling of matter particle velocities through transfer functions, and in doing so examine to what degree the decrease in correlation with initial conditions may be contaminated by velocity-based nonlinearities. With a thorough analysis of the monopole-quadrupole ratio, we find the best-fitting values for the Eulerian velocity dispersion, $\sigma_p = 378.3$ km/s { for a Lorentzian finger-of-God damping factor and $\sigma_p = 254.6$ km/s for a Gaussian one}. The covariance of the cosmological linear growth rate $f$, is estimated in the Eulerian and Lagrangian cases. Comparing Lagrangian and Eulerian, we find that the error in $f$ improves by a factor of 3, without the need for nonlinear velocity dispersion modelling. 
\end{abstract}

\maketitle


\section{\label{sec:intro} Introduction}

The large-scale structure of the matter distribution of the universe is a cosmological observable that is the subject of much observational optimism. The three-dimensional information contained in the positions of galaxies that trace a density distribution can provide measurements of both standard $\Lambda$CDM parameters, and provide constraints on more exotic models for dark matter. This has led to an outpouring of interest in ever larger scale surveys, for instance, the recent successes of the Sloan Digital Sky Survey, and upcoming projects such as the Legacy Survey of Space and Time, and the Dark Energy Spectroscopic Instrument  \cite{SDSS16, LSSTWhitePaper, DesiWhitePaper}.  

However, measurements of matter made from galaxies on the sky are exclusively made in redshift space, which couples both the cosmological information and the astrophysical peculiar velocities.

The problem of determining the position of astrophysical tracers based on their redshift is simple to propose, yet in practice difficult to unravel. Astrophysical objects are subject to nonlinear dynamics that are independent from the Hubble expansion. That said, the regimes where linear dynamics apply can be a rich source of cosmological information, measuring the linear growth rate of the universe \cite{Kaiser1984,RSDPercivalWhite}. Much of this information is encoded in the mapping from redshift space to real space. One of the key connection between real space and redshift space is our ability to model and understand the velocity structure of the large scale density field \cite{Sailer2021}.  

Velocity as a vector quantity can be assigned a one dimensional power spectrum by computing the power spectrum in each direction of the velocity and summing over the result. The formalism of linear transfer functions provides a method to extract a velocity field from the real space density field of dark matter \cite{InmanVelocity}. However, a measurement of the scale at which the linear theory of transfer functions fails to accurately reproduce the underlying velocity field has not been concretely connected to our ability to predict the power spectrum in redshift space. Further, recent interest in recovery of the linear density field through Lagrangian reconstruction algorithms especially recent work in the context of higher-order Lagrangian perturbation theory, e.g., Refs.\cite{chen2021, vlah2015, hivon1995} motivates an analysis of how well the density field in \textit{Lagrangian} space is able to reproduce the velocity structure via transfer functions \cite{EmodeYuPenZhu, ZhuNonlinearRecon}. 

This paper addresses the regimes for which theoretical first-order models of redshift space distortions reproduce the full nonlinear results at the velocity level. In Section \ref{sec:theory}, we recall in detail the linear theory picture of redshift space, and equivalent notions in Lagrangian space. In Section \ref{sec:results}, we present results from the CUBE $N$-body simulation \cite{CUBEPaper}, comparing the correlations for the density field power spectrum in Eulerian and Lagrangian space, and the velocity power spectrum. We apply these results to a nonlinear fitting form for the redshift-real space mapping in Section \ref{sec:nonlinear}, and perform a covariance analysis for extracting the linear growth rate $f$ from the monopole-quadrupole ratio. We discuss implications and conclude in Section \ref{sec:discuss}.

\section{Theory\label{sec:theory}}

\subsection{Linear Theory and Kaiser Approximation}

In the regime of linear perturbation theory with standard isotropic and homogenous assumptions, we can relate the overdensity field at redshift $z$, $\delta(\bm{k},z)$ to the initial overdensity field $\delta_{\text{ini}}(\bm{k})$ using the transfer function in Fourier space. The following equations: 

\begin{equation}\label{eqn:linper}
    \delta(\bm{k}, z) = T_\delta(k,z)\delta_{\text{ini}}(\bm{k}),
\end{equation}

\begin{equation}\label{eqn:linpervel}
    \bm{v}(\bm{k},z)=T_{\bm{v}}(k, z) \delta_{\text{ini}}(\bm{k}) \bm{\hat{k}},
\end{equation}

describe this relationship. In the above, $T_\delta(k,z)$ is the matter transfer function at redshift $z$, and $T_{\bm{v}}(k, z)$ is the velocity transfer function at redshift $z$. Matter transfer functions can be calculated from theoretical considerations, as derived from the initial conditions and the curvature perturbations \cite{MoWhite}. A standard linear transfer function relates the gravitational potential $\phi$ to the growth factor $D(t)$ and scale factor $a(t)$ in the following form:

\begin{equation}\label{eqn:transferfn}
T_\delta(k,t_m) = \kappa \beta(k) \frac{\phi(k, t)}{\beta(k)}\frac{D(t_m)}{a(t_m)}\frac{a(t)}{D(t)}.
\end{equation}

In the above equation, $\beta(k)$ encodes isocurvature or isentropic perturbations to the spacetime metric, $\kappa$ is a normalization constant, and $t_m$ refers to the time at which the universe enters an the Einstein-de Sitter phase \cite{MoWhite}. The redshift corresponding to $t_m$ is $z_m=10$ \cite{MoWhite}. For linear prescriptions of $\phi(k,t)$, it is possible to separate Equation \ref{eqn:transferfn} into a scale dependent transfer function $T_\delta(k,t)= D^2(t)T_\delta(k)$.

For this work we used the CAMB code to calculate the scale dependent matter transfer function $T_\delta(k)$ \cite{CambCode}. From the matter transfer function it is possible to calculate an approximate velocity transfer function by way of the continuity equation \cite{InmanVelocity}. With these transfer functions provided, we can extract a velocity potential $\phi_{\bm{v}}(\bm{k})$ from the nonlinear $\delta(k,z)$:

\begin{equation}\label{eqn:linvel}
    \bm{v}(\bm{k},z)=i \bm{k} \phi_{\bm{v}}(\bm{k})= \frac{T_{\bm{v}}(k,z)}{T_\delta(k,z)}\delta(\bm{k},z) \frac{\bm{k}}{|k|}.
\end{equation}

An inverse Fourier transform of this calculated quantity gives us the real velocity field. This gives us a prescription for calculating the velocity field based on the current observed density field. 

Turning to redshift space, we define the relationship from real space (indexed by $\mathbf{r}$) to redshift space (indexed by $\mathbf{s}$)  adds the velocity $v_{||}$ along the line of sight direction $\hat{\mathbf{r}}_{||}$ by the mapping:

\begin{equation}\label{eqn:zshift}
    \mathbf{s}= \mathbf{r}+ v_{||} \hat{\mathbf{r}}_{||}.
\end{equation}

 The most common simplification applied to this mapping is the Kaiser line-of-sight approximation, from which arises the Kaiser formula: 

\begin{equation}\label{eqn:kaiser}
    \delta^s(\bm{k}, \mu)= \left(1+f \mu^2\right) \delta(\bm{k}).
\end{equation}

Here, $\mu= k_{||}/k$ is the angle along the line of sight and $f= \Omega_m^{6/11}$. We will use Equation \ref{eqn:kaiser} as our basis for the linear redshift space density power spectrum, and Equation \ref{eqn:linpervel} as our basis for the linear velocity power spectrum contributions. 

\subsection{Lagrangian Space}

The theory of this work touches on Lagrangian space viewpoints of the matter density field. In this section we provide a brief summary of this formalism. 
The most common view of the cosmological fluid is to model it as a grid of densities and velocities, where each cell's observables are associated with the particles that occupy it at a given time. This Eulerian view can be contrasted with the Lagrangian view, where individual particles are followed, and their movements tracked by a displacement field. Here we summarize the Lagrangian picture of dark matter particles and its connection to velocity fields. The coordinate $\textbf{x}$ of a particle in the Eulerian grid is connected to the initial Lagrangian coordinate $\textbf{q}$ by the \textit{displacement field} $\mathbf{\Psi}(\textbf{q})$ \cite{LPTResimJenkins}:

\begin{equation}\label{eqn:xtoq}
    \mathbf{x}(\textbf{q})= \textbf{q}+ \mathbf{\Psi}(\textbf{q}).
\end{equation}

Taking the linear approximation to the full fluid equations, we relate the displacement field to the gravitational potential $\phi(\bm{k})$:

\begin{equation} \label{eqn:lptpsi}
\mathbf{\Psi}(t,\bm{k}) = -D(t) \nabla_q \phi(\bm{k}).
\end{equation}

Here, $\nabla_q$ refers to the gradient operator with respect to the Lagrangian coordinate. Naturally, these potentials are connected to the overdensity field by a Poisson equation, which taking the divergence of Equation \ref{eqn:lptpsi}, leads us to:

\begin{equation} \label{eqn:lptpsiden}
-\nabla \cdot \mathbf{\Psi} \propto  \delta(\textbf{q}).
\end{equation}

Here, proportionality is contained in the growth factor $D(t)$. This theory is additionally used to establish initial conditions in most $N$-body simulations, occasionally going to second order which has similar structure \cite{LPTResimJenkins}. For the remainder of the paper, we will refer to the Lagrangian space overdensity field as $\delta_{\text{Q}} = -\nabla \cdot \Psi$, and the Eulerian overdensity as $\delta_{\text{E}}$. Finally, we define the equivalent redshift space quantities in Lagrangian space, for the displacement field:

\begin{equation}\label{eqn:psis}
   \mathbf{\Psi}^s(\textbf{q}) =  \mathbf{s}(\textbf{q})- \textbf{q},
\end{equation}
and for the overdensity
\begin{equation}\label{eqn:lagdelta}
  \delta^s_{\text{Q}} = -\nabla \cdot \Psi^s,
\end{equation}

which allows us to proceed to comparing them fairly to their Eulerian space equivalents.

\begin{figure*}
  \centering
  \includegraphics[width=\linewidth]{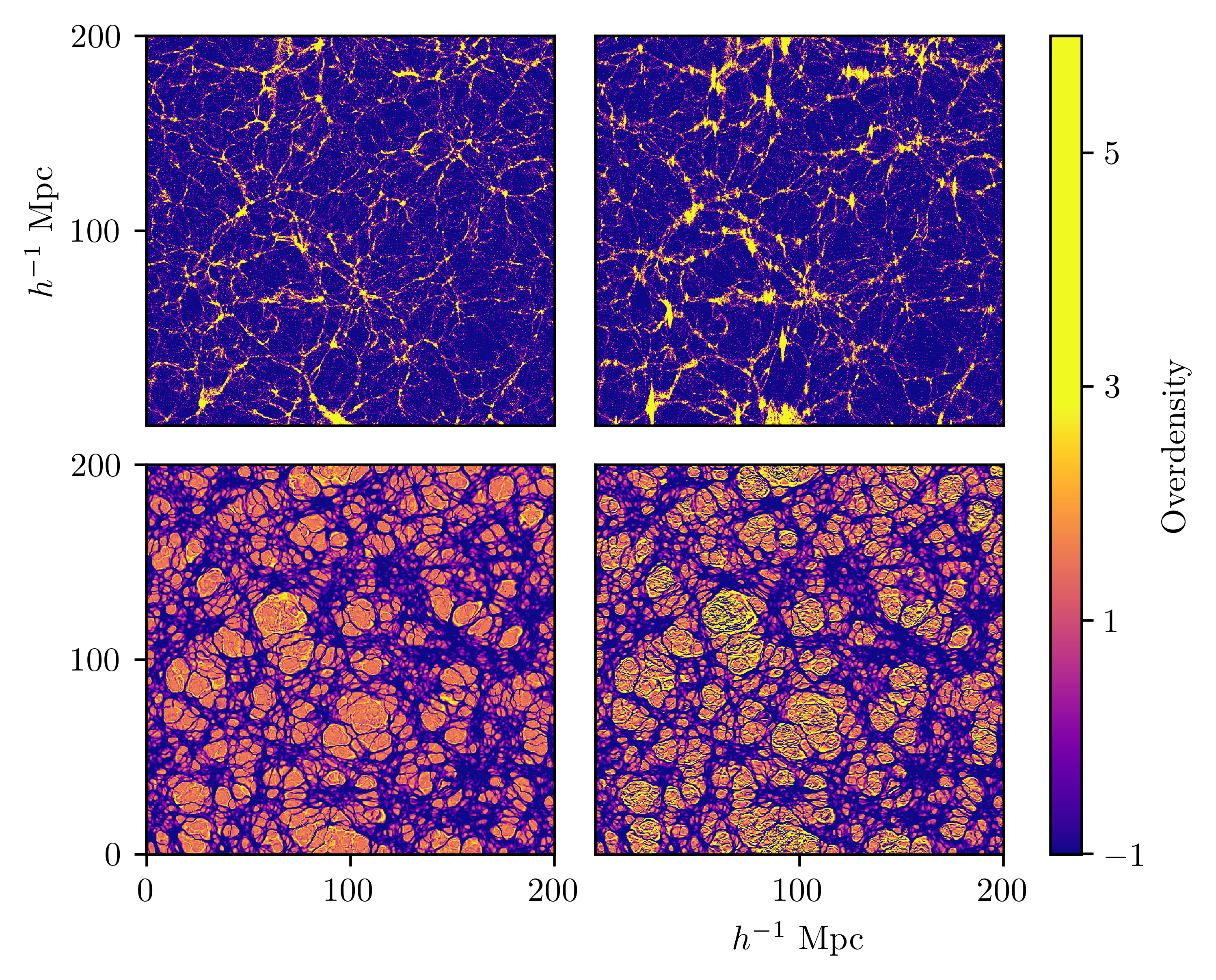}
  \caption{Slices of the CUBE simulation, with width $200 h^{-1}$ Mpc and thickness $4.0$ $h^{-1}$ Mpc, at $z=0$. The field is constructed using a cloud-in-cell (CIC) interpolation scheme. (Top left) The Eulerian nonlinear density field. (Top right) The Eulerian nonlinear density field in redshift space. (Bottom left) The Lagrangian nonlinear density field $\delta_{\text{Q}} = -\nabla \cdot \Psi$. (Bottom right) The nonlinear density field calculated according to Lagrangian theory, in redshift space $\delta^s_{\text{Q}} = -\nabla \cdot \Psi_{s}$.}
  \label{fig:DensitySlices}
\end{figure*}

\section{Results \label{sec:results}}

In this section we detail the results of the CUBE $N$-body simulation, analyzing the real and redshift space matter power spectra, their velocity power spectra, and the redshift space quadrupole-monopole moment for the purposes of fitting the linear growth parameter $f$. 

The CUBE simulation was run with $N_{\text{cdm}}= 800^3$ particles in a $L=200 h^{-1}$ Mpc length box \cite{CUBEPaper}. The cosmological parameters used are: $\Omega_m=0.29, h = 0.72, \sigma_8 = 0.8, n_s = 0.8$, as given by the Dark Energy Survey parameters \cite{DESParams}. The initial conditions of the dark matter simulation are provided by a Gaussian noise map, multiplied by a transfer function to the starting redshift $z_{\text{ini}}=100$. The scale-dependent transfer function is generated by the CAMB code \cite{CambCode}, as input to CUBE\cite{CUBEPaper}. Baryonic effects are provided by these transfer functions, for which we used $\Omega_{\text{b}}= 0.044$ \cite{DESParams}. 

\subsection{Density}

The Eulerian density field is calculated using a cloud-in-cell interpolation method. The Lagrangian displacement is calculated by a nearest grid point (NGP) interpolation of the Lagrangian displacement vector calculated for each individual particle ($\mathbf{\Psi} = \mathbf{x}-\mathbf{q}$). The initial Lagrangian positions of the particles are generated from the initial conditions of CUBE, recorded as a particle ID \cite{CUBEPaper}. The density field from linear theory is calculated using Equation \ref{eqn:lptpsiden}. The Lagrangian density field is calculated according to Equation \ref{eqn:lptpsiden} in Fourier space. 

The density field as determined by the dark matter particles in the CUBE simulation is shown in Figure \ref{fig:DensitySlices}, at $z=0$. The finger-of-God effect between the real space density field and the redshift space density field is clearly visible in the elongation along the line of sight axis. We note that visually, the density field in real and redshift space calculated by Lagrangian space methods shows greater similarity to a Gaussian random field than their Eulerian counterparts.

\begin{figure}[h!]
  \centering
  \includegraphics[width=\linewidth]{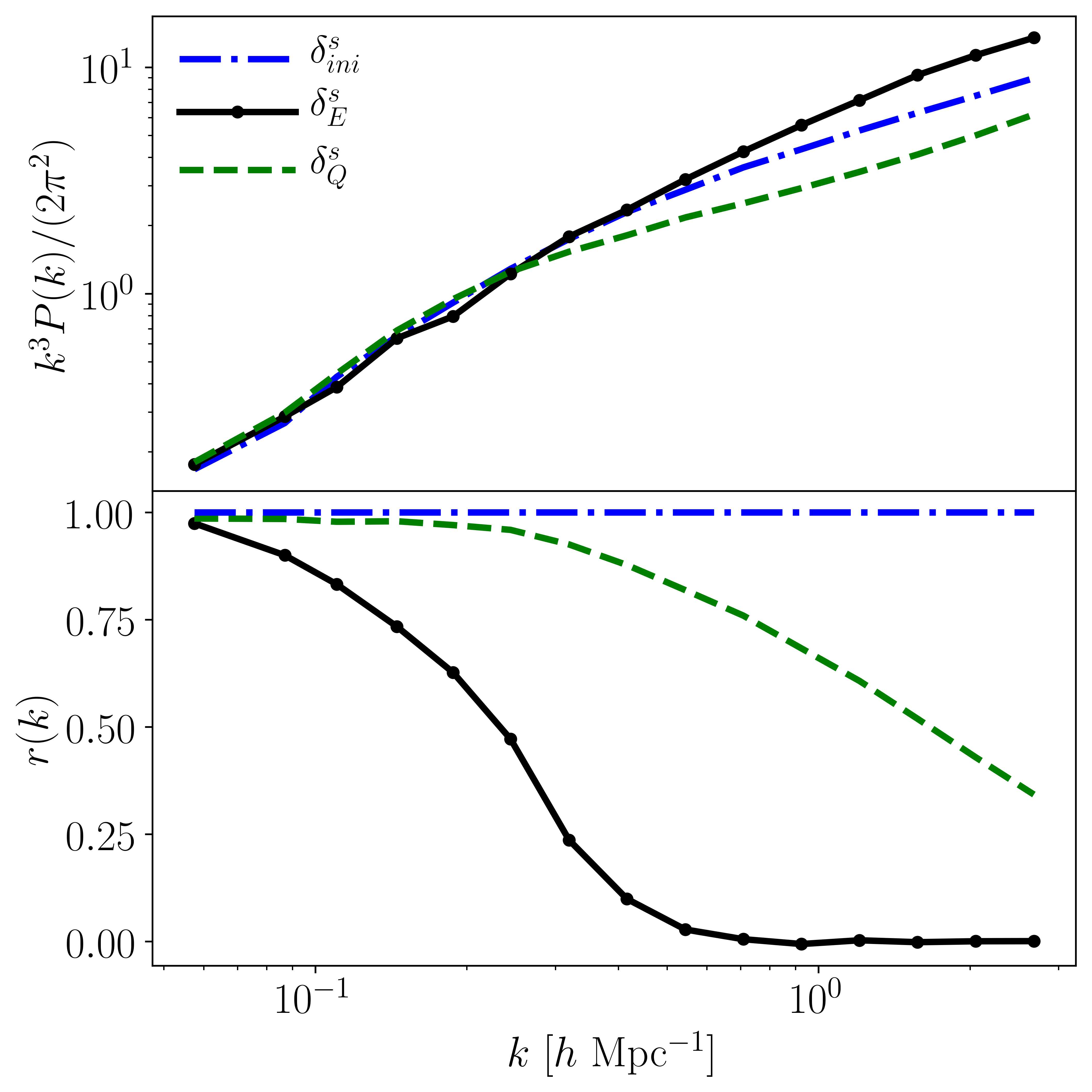}
  \caption{(Top) The dimensionless redshift space matter power spectra for the simulation at $z=0$, compared to the linear theory from initial conditions. The blue dot-dashed line is the linear matter power spectrum. The black dotted line connected by solid lines is the Eulerian matter power spectrum. The dashed green line is the Lagrangian matter power spectrum. As in previous research, the linear power spectrum is well-aligned with the nonlinear power spectrum to $0.1$ $h$ Mpc$^{-1}$.  (Bottom) The correlation coefficient for each of the components with the linear density field. }
  \label{fig:MatterPower}
\end{figure}

The subsequent Figure \ref{fig:MatterPower} shows the matter power spectra for each of the fields, compared to the linear theory counterpart. As was found in multiple computational studies \cite{jalilv2019nonlinear,ZhuRSD}, linear theory ceases to be effective at ~ $0.1 $ $h^{-1}$ Mpc. 

We assess the correlations to linear initial conditions using the standard correlation coefficient:

\begin{equation}\label{eqn:rdef}
    r_{ij}(k) = \frac{P_{ij}(k)}{\sqrt{P_i(k) P_j(k)}},
\end{equation}

where $P_{ij}(k)$ is the cross-power for species $i,j$, and $P_i(k)$ the corresponding auto-power. The correlation coefficient as a function of scale is shown in Figure \ref{fig:MatterPower}. All correlations are with the linear initial conditions provided to CUBE. For the  Eulerian redshift space density field, the correlation is above 0.9 for $k<  0.09$ $h$ Mpc$^{-1}$ and above 0.75 for $k< 0.1$ $h$ Mpc$^{-1}$. For the Lagrangian redshift space density field, we see an improvement to above 0.90 for $k < 0.4$ $h$ Mpc$^{-1}$ and above 0.75 for $k< 0.6$ $h$ Mpc$^{-1}$. We use this to benchmark the contributions to nonlinearities from the velocities themselves.

\subsection{Velocity}

The velocity fields for the simulation are generated by an averaged velocity field with a residual component calculated from the integrated particles of the $N$-body simulation. The smoothing scale over which this averaging is accurate can be approximated by the velocity variance, given by $\sigma_v = 5.62 $km/s in our simulation. Details of the velocity storage for CUBE are given in \cite{CUBEPaper}.

\begin{figure*}
  \centering
  \includegraphics[width=0.98\linewidth]{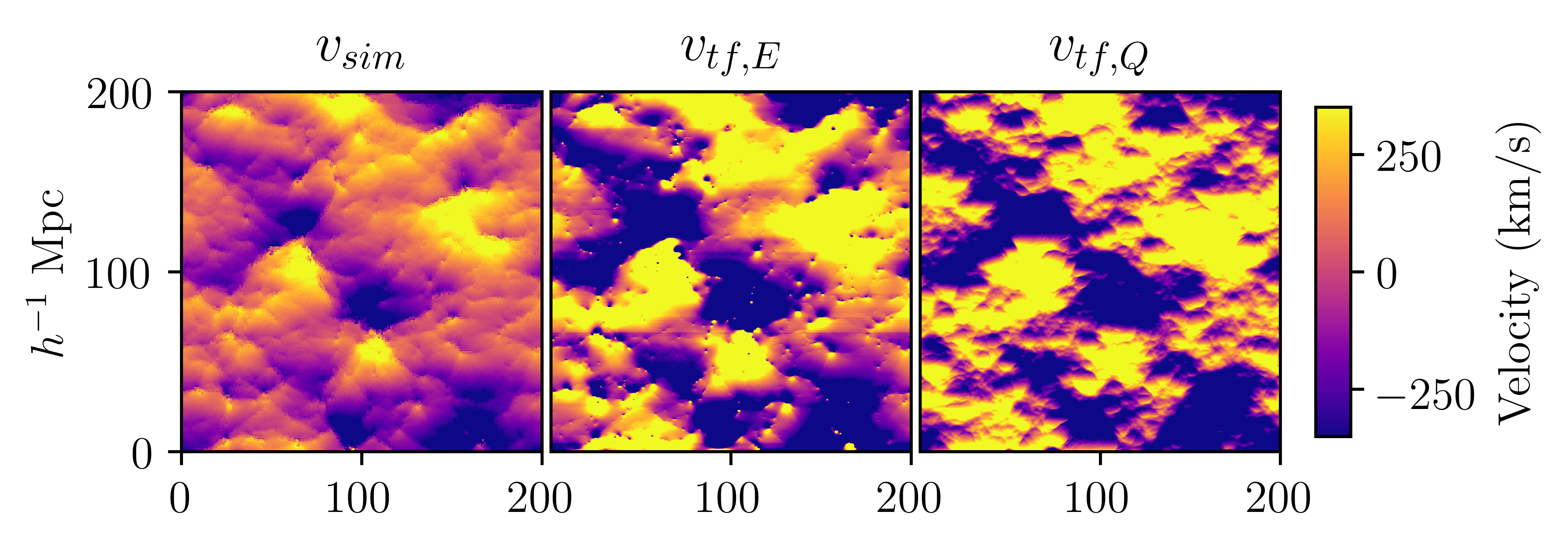}
  \caption{Slices of the CUBE simulation, with width $200$ $h^{-1}$ Mpc and thickness $4.0$ $h^{-1}$ Mpc, at $z=0$ (redshift). (Left) The line-of-sight velocity field calculated from an nearest grid point interpolation of the particles and historical particle velocities ($v_{{sim}}$). (Middle) The line-of-sight velocity field calculated applying transfer functions to the Eulerian density field, according to Equation \ref{eqn:linpervel} ($v_{{tf, E}}$). (Right) Same as the middle panel, but using the Lagrangian density field ($v_{{tf, Q}}$). For all panels, the line-of-sight direction is taken to be the $\hat{x}_2$ direction of the simulation grid, with positive velocities moving away from the viewer.}
  \label{fig:VelocitySlices}
\end{figure*}

Figure \ref{fig:VelocitySlices} shows velocity slices from the simulation, either derived directly from the particle positions, or from the density fields (Eulerian, Lagrangian) using transfer functions. In these slices we see qualitative agreement of the velocity structure, especially at large scales. The Eulerian density field shows most significant visual agreements to the true velocity field.

\begin{figure}[htp]
\centering
  \includegraphics[width=\linewidth]{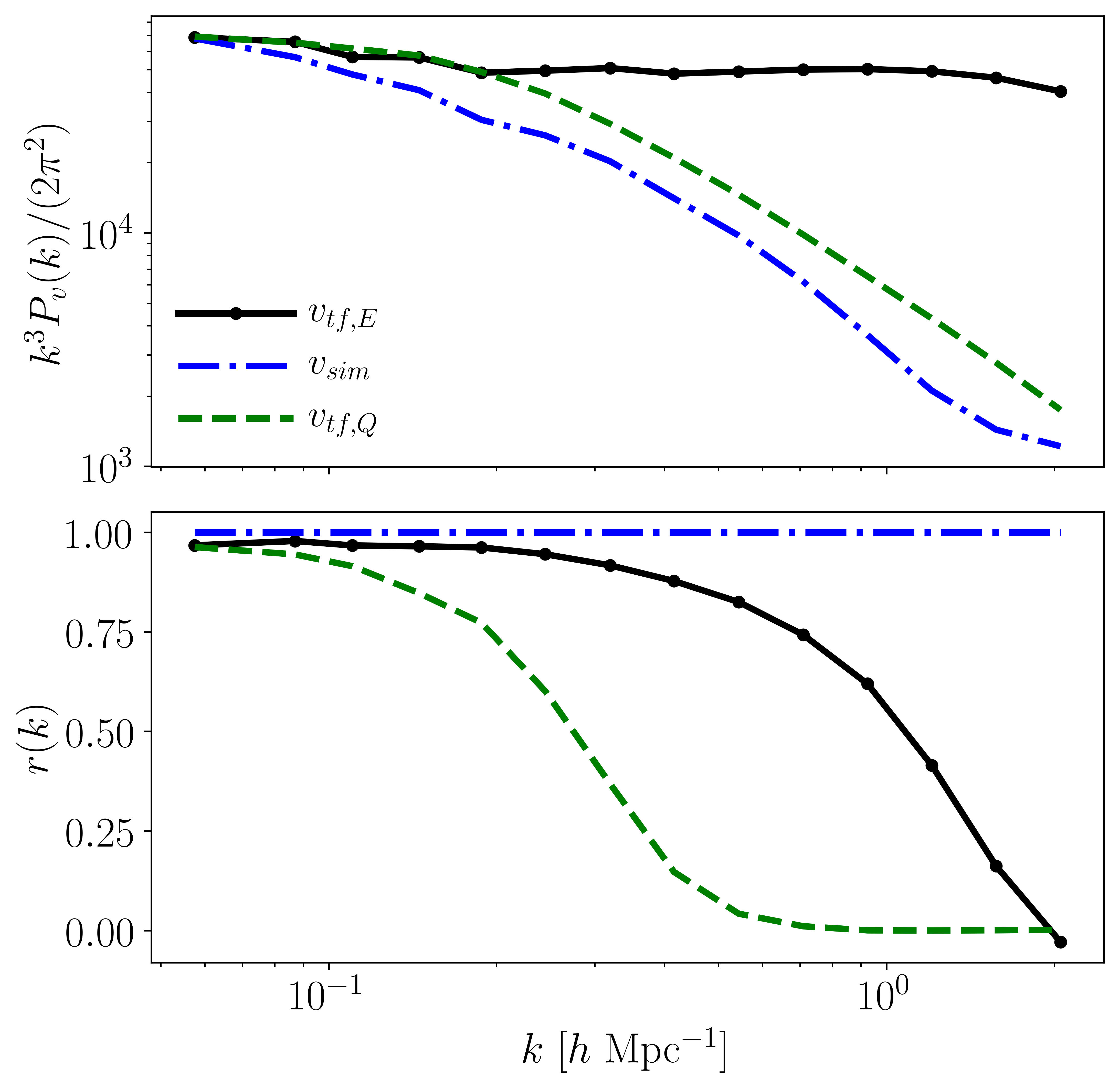}
  \caption{The dimensionless power spectrum of the velocity according to Equation \ref{eqn:powerlinvel}. The blue dot dashed line shows the true simulation velocity power spectrum. The black dotted line connected by solid lines is the velocity power spectrum of the density-derived velocity field in Eulerian space. The green dashed line shows the velocity power spectrum of the density-derived velocity field in Lagrangian space.}
  \label{fig:VelocityPower}
\end{figure}

The power spectrum for velocity, defined by the sum of the power spectra over all directions, is shown in Figure \ref{fig:VelocityPower}. According to Equation \ref{eqn:linvel}, we find the linear expression for the power spectrum transfer function prescription to be

\begin{equation}\label{eqn:powerlinvel}
    P_{v,tf}(k,z)= \frac{T^2_v(k,z)}{T^2_\delta(k,z)}P(k,z) .
\end{equation}

If we compare the velocity derived from the Eulerian density field to the true simulation velocity, the power of the true simulation velocity drops off much faster at smaller scales. This is consistent with the concept of small nonlinear motion as modelled by Finger-of-God effects. 

The correlation coefficient as a function of scale is shown in Figure \ref{fig:VelocityPower}. In this case the correlation coefficient refers to the cross-power with the true simulation velocity at $z=0$, to assess how well linear theory with transfer functions can predict the velocity power spectrum.  For the nonlinear density field with transfer functions applied, the correlation coefficient is greater than 0.9 for $k<  0.2$ $h$ Mpc$^{-1}$ and above 0.75 for $k< 0.3$ $h$ Mpc$^{-1}$. For the Lagrangian density field with transfer functions applied, the performance worsens with a steep drop, about 0.90 for $k < 0.1$ $h$ Mpc$^{-1}$ and above 0.75 for $k< 0.13$ $h$ Mpc$^{-1}$.

\subsection{Error Estimates}
\label{sec:errorestimates}
In order to better assess how these correlations to linear theory best translate to accurate estimates of cosmological parameters, we need to establish a scheme for error estimation. The Lagrangian space power spectrum is known to be subject to Gaussian errors \cite{FisherInman}. The Lagrangian monopole and quadrupole have analytical (Gaussian) errors, given in the form:

\begin{equation}\label{eqn:gaussdiag}
C^{Gauss}(k_i, k_j) = \frac{2}{N(k)} (P^{S}(k) + P^{N}(k))^2 \delta^D_{k_i k_j}, 
\end{equation}

where $N(k)$ is the number of samples per $k-$bin, $P^{S}(k)$ refers to a signal-component contribution, $P^{N}(k)$ to a noise-component contribution, and $\delta^D_{k_i k_j}$ is the Dirac-delta function. 

\begin{figure}[htp]
\centering
  \includegraphics[width=\linewidth]{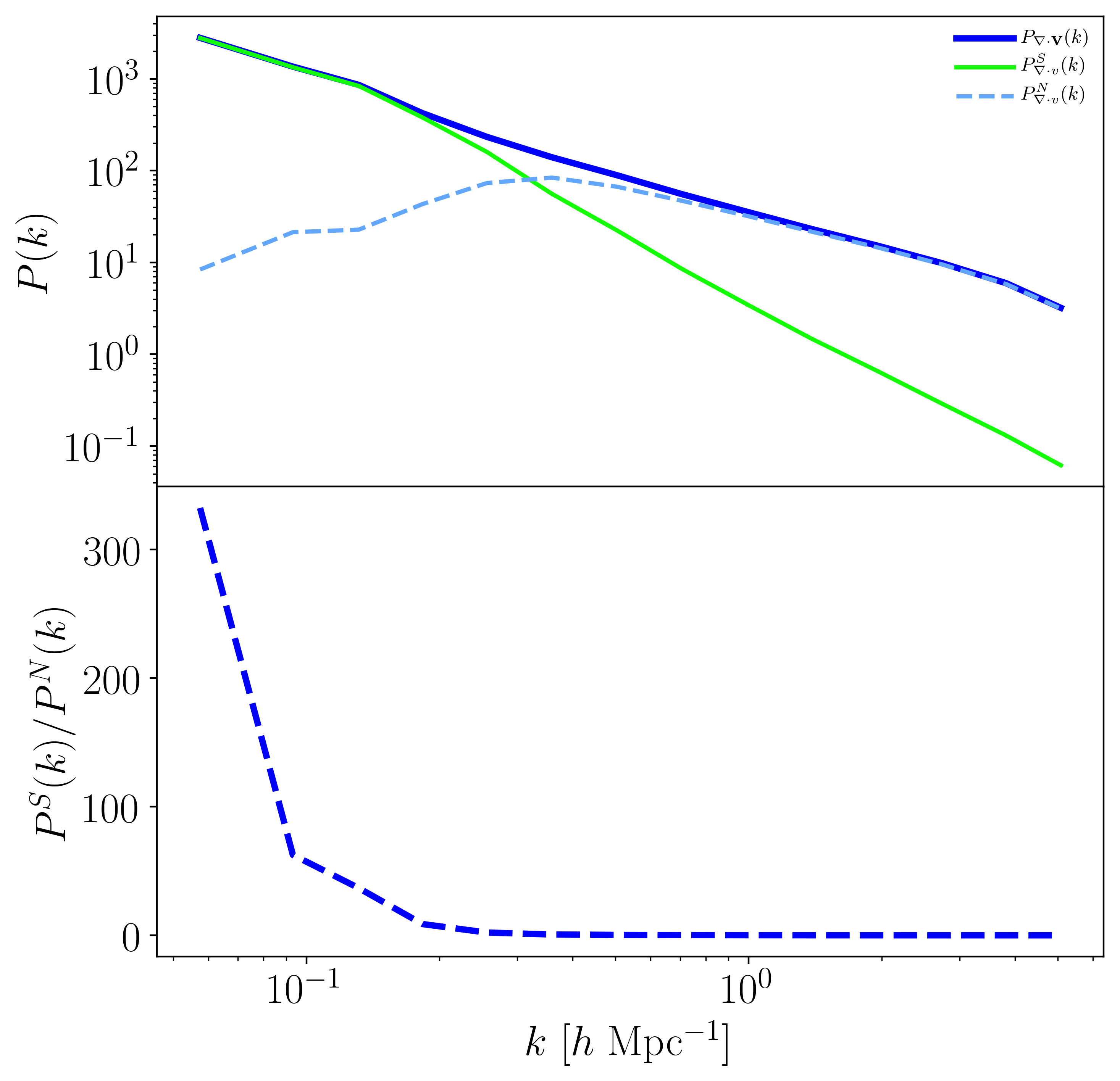}
  \caption{(Top) Plots of the velocity divergence power spectrum and its decomposition. The noise component of this quantity quantifies the additional error in the Lagrangian redshift space power spectrum (above the Gaussian prescription). (Bottom) The signal-to-noise ratio as defined by Equation \ref{eqn:noiseeqn}.}
  \label{fig:velcorr}
\end{figure}

In the Lagrangian case, the contributions to the redshift space matter power spectrum from velocities contain a component that is coupled with the linear theory (per Equation \ref{eqn:linvel}), and a component that includes nonlinear random motions \cite{ScoccimarroRSD}. Therefore, we separate signal (correlated with the real space Lagrangian power) and noise (completely uncorrelated) components of the velocity divergence using the correlation coefficient according to:

\begin{equation}\label{eqn:noiseeqn}
  P^N(k) = (1-r(k)^2)P_{\nabla \cdot \mathbf{v}}(k),  
\end{equation}

where $r(k)$ is the correlation coefficient between $\nabla \cdot \mathbf{v}$ and $\delta_{Q}$, and $P_{\nabla \cdot \mathbf{v}}(k)$ is the power spectrum of the velocity divergence. Figure \ref{fig:velcorr} shows the contribution of signal and noise to the velocity divergence at different scales. Here $P^S(k)$ refers to the nonlinear Lagrangian real space power spectrum.

The error in the Eulerian space power spectrum has a more complex nonlinear expression sourced from \cite{JoachimErrors}. The procedure is briefly outlined in Appendix \ref{appn:errors}. 

\section{Nonlinear Modelling \label{sec:nonlinear}}

\begin{figure}[htp]
\centering
  \includegraphics[width=\linewidth]{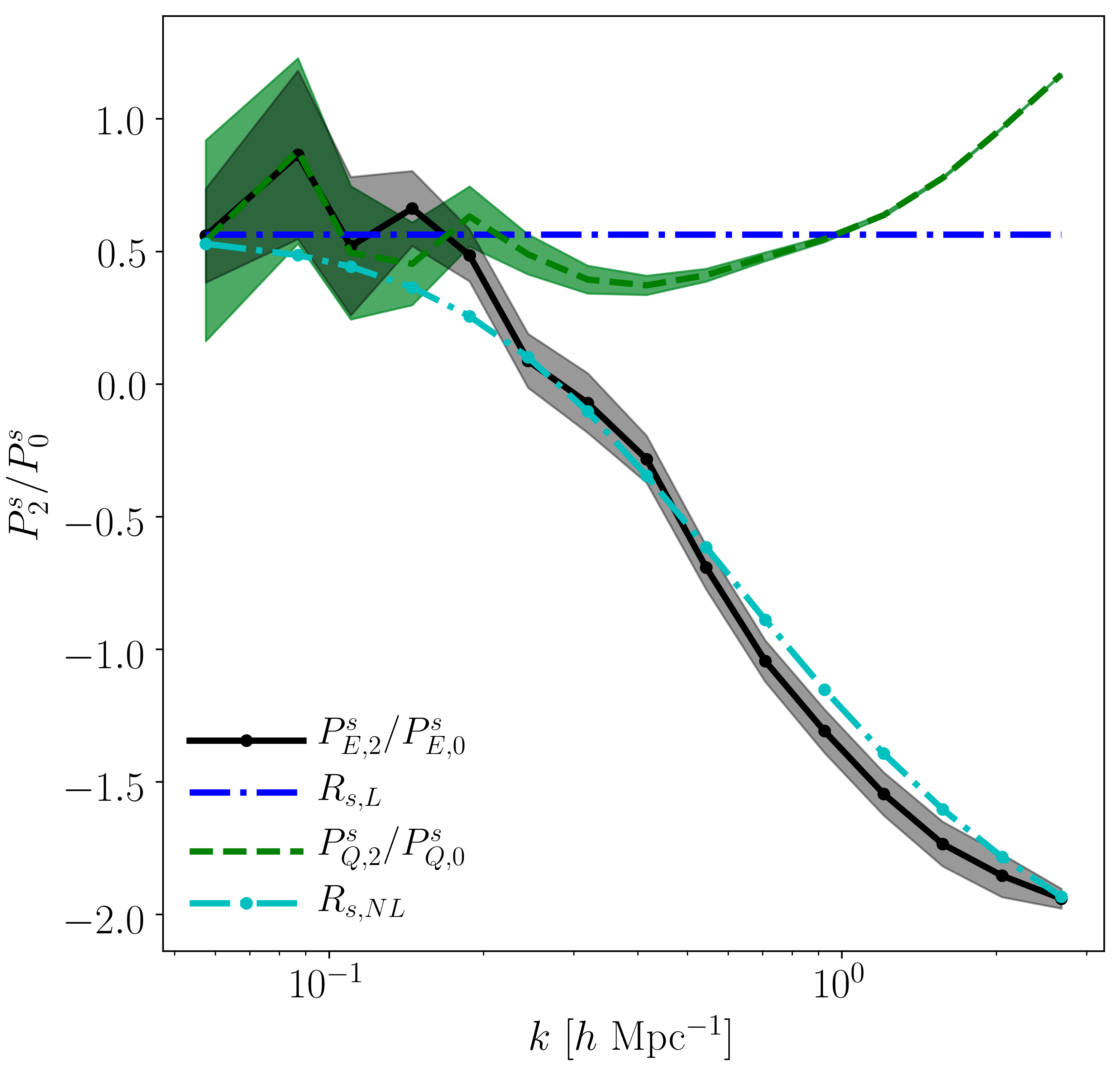}
  \caption{The ratio of the quadrupole to monopole moment. The spikes at low $k$ are attributed to low statistics contributing to the quadrupole moment. The blue dotted line connected by solid lines shows the linear quadrupole-monopole ratio. The black dotted line connected by solid lines is the Eulerian nonlinear quadrupole-monopole ratio. The cyan dot-dashed line is the best fit for the predicted quadrupole-monopople ratio given by Equation \ref{eqn:fingerofgod}. The parameter $\sigma_p = 378.3$ km/s was found to be the best fit for Equation \ref{eqn:fingerofgod}. Note the substantial shrinking of error bars in the Lagrangian case (green region) when compared to the Eulerian case (grey region). This provides a strong case for investigation into Lagrangian space methods to take full advantage of redshift space distortions as a cosmological probe.} 
  \label{fig:EulerianFit}
\end{figure}

For a direct application of the correlations outlined in Section \ref{sec:results}, we turn to using ratios of multipole moments for fitting cosmological parameters. In the general case:

\begin{equation}\label{eqn:ratiogeneral}
    R_s({k}) = \frac{P^s_2({k})}{P^s_0({k})}= \frac{\int_{-1}^{1}(3 \mu^2 - 1)P^s(k,\mu) d\mu}{\int_{-1}^{1}P^s(k,\mu) d\mu}.
\end{equation}

In the above equation, $P_2(\mathbf{k}), P_0(\mathbf{k})$ refer to the quadrupole and monopole moments for a given redshift space power spectrum (Lagrangian, Eulerian). In the second equality, we assume the line-of-sight approximation (redshift space distortions only along the line of sight $\mu$). As in previous discussions, we will use the linear prescription, Equation \ref{eqn:linper}, in order to fit the Lagrangian quadrupole-monopole ratio. Due to the persistent presence of nonlinearities in the small-scale regime, the Eulerian quadrupole-monopole ratio displays a decrease in this regime. Therefore, for the Eulerian quadrupole-monopole ratio, we consider a common nonlinear mapping of redshift space distortions that includes the Fingers-of-God effect \cite{WhiteRedshift2015}. The following formula proposes an exponential dispersion relation in real space, which in Fourier space is given by \cite{WhiteRedshift2015}:

\begin{equation}\label{eqn:fingerofgod}
    P_{s,\text{NL}}(k, \mu)= (1+f \mu^2)^2\frac{ P_{ini}(k)}{1+k^2 \mu^2\sigma_p^2/2}.
\end{equation}

Here, $P_{ini}(k)$ refers to a linear prescription for the initial power spectrum, and we use $ P_{s,\text{NL}}$ to label the fit to the Eulerian data, rather than the Eulerian redshift space power spectrum sourced from $\delta_E$. Historically the parameter $\sigma_p$ has been numerically determined by fits to simulations \cite{WhiteRedshift2015}, \cite{GrasshornSpherical2020}, \cite{ScoccimarroRSD}, \cite{TaylorMultipole1996}. It should be cautioned that the mathematics of this dispersion $\sigma_p$ do not correspond directly to a dispersion associated with a velocity probability distribution function (see \cite{ScoccimarroRSD} for a full analysis). Appendix \ref{appn:integrate} shows the full integrated statements of Equation \ref{eqn:ratiogeneral}, both the linear fit $R_{s,L}$ and the nonlinear fit $R_{s,NL}$. The latter requires us to fit for the parameter $\sigma_p$, a standard technique described in detail in Taylor et. al. \cite{TaylorMultipole1996}. This additional parameter and prescription does not apply to the Lagrangian quadrupole-monopole ratio, as it does not experience the same decay at small scales.

In order to account for uncertainty at different $k$-values, we examine individual fits to each wavenumber $k$. Since the inversion of Equation \ref{eqn:fingerofgod} for the integrated $P_s(k) = \int_{-1}^{1} P_s(k,\mu) d\mu$ is highly nonlinear, we use a root-finding algorithm to determine $\sigma_p$. This resulted in a fit of $\sigma_p = 378.3$ km/s for the Eulerian space power spectrum. Figure \ref{fig:EulerianFit} shows the success of the fit parameters, with the uncertainty regions derived from the diagonal of the covariance matrix. 

Finally, we assess the covariance assigned to the model parameter of interest, the linear growth rate $f$, using the Fisher matrix formalism. Treating $\sigma_p$ as a fixed parameter extracted from simulations, rather than one that is fitted for empirically, the Fisher matrix $\mathcal{F}$ becomes a single number that allows us to calculate the covariance $\Delta f$. We use the following treatment to determine the error in $f$:

\begin{equation}\label{eqn:fisher}
   \Delta f (k_{\text{max}}) = \sqrt{\frac{1}{\mathcal{F}}} = \left(\sum_{k_i, k_j} \left(F(k_i,k_j)\right)^2 \left(\frac{\partial R_s}{\partial f}\right)^2 \right)^{-1}.
\end{equation}

Here $F(k_i,k_j) = C^{-1}(k_i,k_j)$, i.e. we are projecting the Fisher matrix onto the parameter space. $R_s$ is the quadrupole-monopole ratio defined in Equation \ref{eqn:ratiogeneral}, with the understanding that the ratio theoretical fit depends on the linear growth rate $f$. 

\begin{figure}[htp]
\centering
  \includegraphics[width=\linewidth]{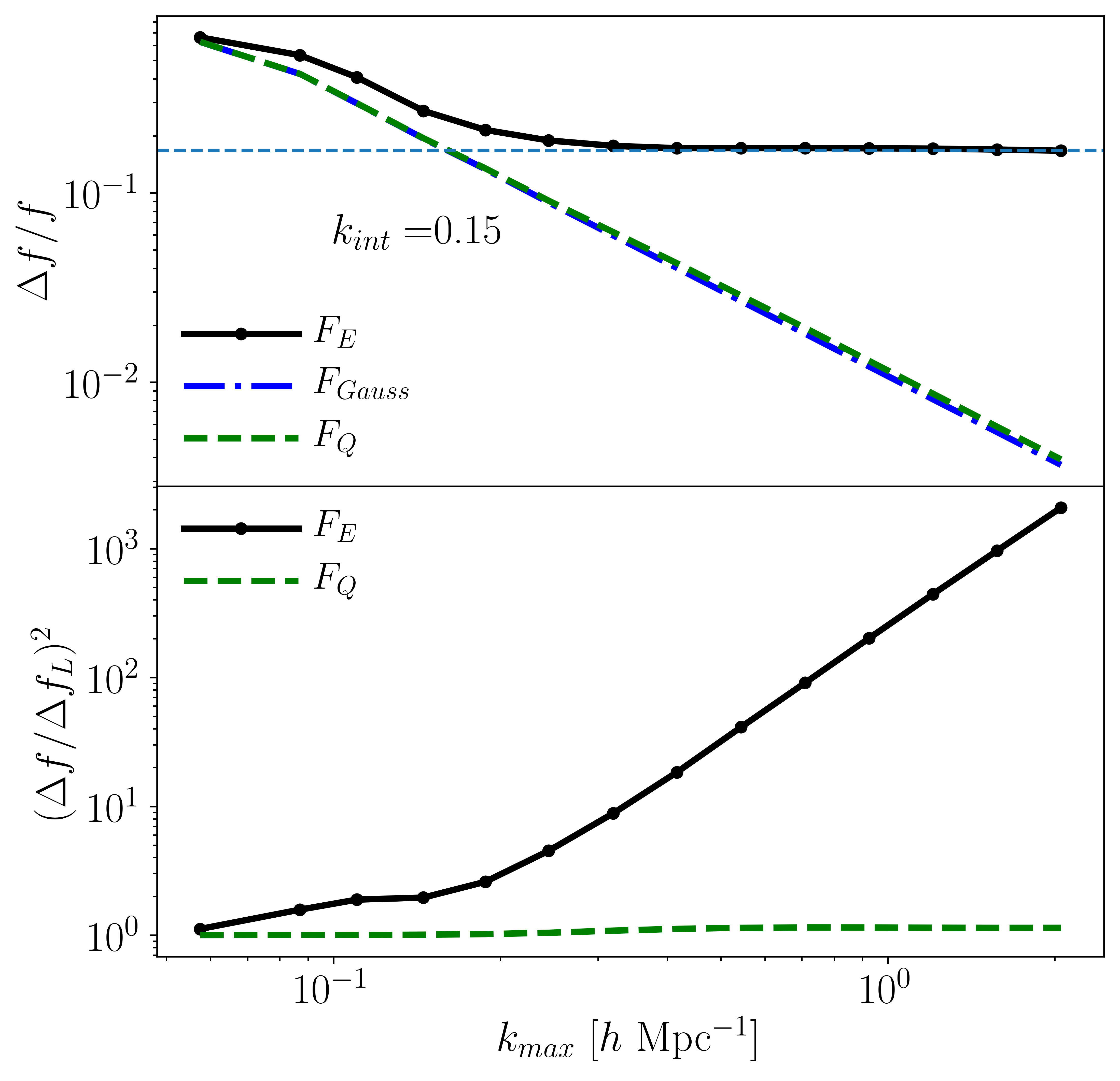}
  \caption{(Top) The parameter covariances for $f$, as determined by Equation \ref{eqn:fisher} and scaled against the simulation value for $f$. We emphasize the Lagrangian covariance's rapid decrease when compared to its Eulerian counterpart. The black dotted line connected by solid lines assumes the fit of Equation \ref{eqn:fullintegratedEuler}, with $\sigma_p = 378.3$ km/s. The green dashed line assumes the fit of \ref{eqn:fullintegratedLinear}, and includes the noise attributed to the Lagrangian power spectrum as described in the appendix. The blue dash-dotted line assumes the the fit of Equation \ref{eqn:kaiser}, and comes from the Gaussian errors described in Sec. \ref{sec:errorestimates}. The light blue dashed line is the asymptotic behavior of the nonlinear Eulerian Fisher estimates, meeting the Lagrangian estimates at $k \approx 0.1$. (Bottom) The square of the parameter covariances in the top panel, this time scaled by the linear signal-only covariance (blue dot-dashed line on the top plot). }
  \label{fig:fisherplot}
\end{figure}

Figure \ref{fig:fisherplot} depicts the parameter covariance as a function of the maximum $k$ included in the fitting procedure. The constraining power on the parameter $f$ shows substantial improvement in the nonlinear range with linear errors applied. With the most conservative estimate of linear correlations persisting to $k \approx 0.3$, the Lagrangian space power spectrum provides  $6$\% errors on the parameter $f$, as opposed to the  $17$\% errors found in the Eulerian case. 

\section{Discussion and Conclusion \label{sec:discuss}}

We have revisited the successes of linear theory at predicting the redshift space power spectrum, and have examined those successes in the context of the velocity prediction from linear theory. We have compared these predictions to their equivalents in Lagrangian space. 

The linear transfer functions used to restore the velocity field at nonlinear scales saw a doubling of the number of highly correlated modes at the Eulerian density field level. The coupled nature of the velocity and density perturbations allows the transfer function formalism, widely derived from density perturbations, to be an accurate predictor of the large-scale velocity.

Advantages gained in Lagrangian space that have been studied at the density field level may in fact be a source of the decrease in velocity correlation in Lagrangian space. The nonlinearities from the Eulerian density field are suppressed in the Lagrangian case due to the Jacobian of the Lagrangian space mapping \cite{EmodeYuPenZhu}. This may then affect its correlation with the \textit{nonlinear} velocity field. The complementarity of these two results has interesting implications for potential applications of novel reconstruction algorithms (for instance \cite{ZhuNonlinearRecon}). 

With respect to forecasting, the plot of Figure \ref{fig:fisherplot} illustrates the promised gains of advanced Lagrangian reconstruction, with even conservative estimates providing below ten percent error in the $k_{max} \approx 0.3 $ regime. This is an improvement over Eulerian modelling methods, for instance, Ref. \cite{jalilv2019nonlinear} cite the best case nonlinear modelling (TNS model) as accurate up to  $k_{max} \approx 0.2 $. In the present work, we have used the best-fitting value of $\sigma_p$ to derive constraints on the growth rate $f$ (Fig. \ref{fig:fisherplot}). Previous approaches using the dispersion method and two other models based on perturbation theory find the  linear growth rate \cite{garcia2020} to be biased by 5-10\% over $10 < r/(h^{-1} {\rm Mpc}) < 50$ at $z < 1$, with the TNS model providing the best constraints. A similar analysis from BOSS ($0.43 < z < 0.7$, Ref. \cite{beutler2014} leads to a 3.1\% uncertainty in the growth parameter over the $5 < r/(h^{-1} {\rm Mpc}) < 100$ range. Linear redshift space distortion modelling has been recently found to be overly optimistic (for instance, \cite{PercivalRSD2021}), but the studies have not yet taken into account the gains available through the Lagrangian space picture, in which it was found, however, that the projected galaxy correlation function does retain some anisotropic effects\cite{nock2010}.  It is our hope that this method will be of interest to investigate applying to improving information that can be extracted from DESI, LSST on the Rubin Observatory, and SDSS. 

\section{Acknowledgements}

We thank Joachim Harnois-Déraps, Derek Inman, Roman Scoccimarro, Ruth Durrer, Minji Oh,
Basundhara Ghosh, Mona Jalilvand and Hao-Ran Yu for helpful discussions and comments on the manuscript.  HP acknowledges support from the Swiss National Science Foundation under Ambizione grant PZ00P2\_179934. Ue-Li Pen receives support from Ontario Research Fund—research Excellence Program (ORF-RE), Natural Sciences and Engineering Research Council of Canada (NSERC) [funding reference number RGPIN-2019-067, CRD 523638-18, 555585-20], Canadian Institute for Advanced Research (CIFAR), Canadian Foundation for Innovation (CFI), the National Science Foundation of China (Grants No. 11929301), Simons Foundation, Thoth Technology Inc, Alexander von Humboldt Foundation, and the Ministry of Science and Technology(MOST) of Taiwan(110-2112-M-001-071-MY3). Computations were performed on the SOSCIP Consortium’s [Blue Gene/Q, Cloud Data Analytics, Agile and/or Large Memory System] computing platform(s). SOSCIP is funded by the Federal Economic Development Agency of Southern Ontario, the Province of Ontario, IBM Canada Ltd., Ontario Centres of Excellence, Mitacs and 15 Ontario academic member institutions.

\bibliography{CosmoRefs.bib}

\appendix

\section{Eulerian Matter Power Spectrum Error Treatment\label{appn:errors}}

The error bars on Figure \ref{fig:EulerianFit} are calculated using a nonlinear estimator for matter power spectra developed in Ref \cite{JoachimErrors}. This research provides a prescription to estimate the full covariance matrix using a decomposition into spherical harmonics. Each spherical harmonic component of the nonlinear covariance matrix is re-scaled by the Gaussian prescription, and broken down into an eigenvalue decomposition \cite{JoachimErrors}. Eigenvectors are given a fitting prescription as a function of scale $k$. For the error bars in Figure \ref{fig:EulerianFit}, the fit from Ref \cite{JoachimErrors} was performed over a specific binning. We rebin the covariance matrix according to the following formula:

\begin{equation}\label{eqn:rebindiag}
    C_{rebin}(k_i, k_j)= \frac{\sum_{k_{i-1} < k_{l} < k_{i}} \sum_{k_{j-1} < k_{m} < k_{j}}C(k_1, k_m)}{[N(k_i, k_j)]^2}
\end{equation}

Here $N(k_i, k_j)$ refers to the number of samples that have been subsumed into the larger $k$-bin size, as compared to the bin selection from Reference \cite{JoachimErrors}. 

\begin{figure}[htp]
\centering
  \includegraphics[width=\linewidth]{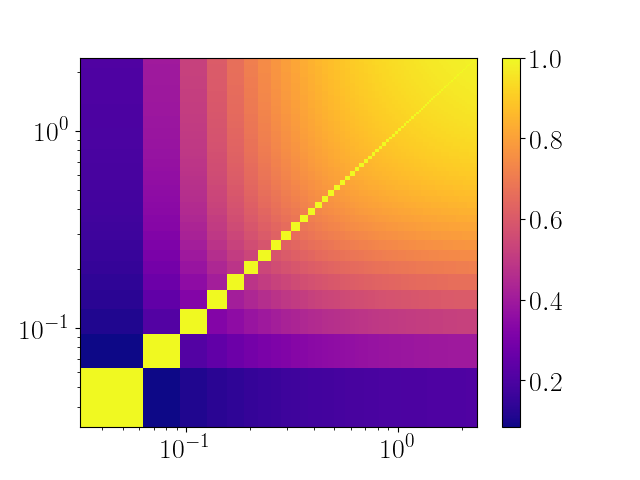}
  \caption{The fitted covariance matrix rescaled by the Gaussian error prescription, with fitting parameters given by \cite{JoachimErrors}. Note that the fit is only valid extending to $k<2.3 h $Mpc$^{-1}$. }
  \label{fig:Covmatrix}
\end{figure}

The fitted covariance matrix, with the diagonal rescaled to unity, is given by Figure \ref{fig:Covmatrix}. As a first pass, we assume that both the redshift space monopole and quadrupole have the same percent error contribution as the contribution given by the diagonal of this fit prescription. 

\section{Integrated Forms of the Multipole Ratio Fit \label{appn:integrate}}

For completeness, we provide the full integrated forms of our prescriptions for the power spectra. First, the linear form dictated by Equation \ref{eqn:linper}, and integrated into Equation \ref{eqn:ratiogeneral}:

\begin{equation}\label{eqn:fullintegratedLinear}
R_{s,L}(f)= \frac{(4/3) f +(4/7) f^2}{1+(2/3) f +(1/5)f^2}.
\end{equation}

Then, we turn our attention to integrating the nonlinear fit to the quadrupole-monopole ratio outlined by Equation \ref{eqn:fingerofgod}. Let $a= {k \sigma_p}/{\sqrt{2}}$. 
\begin{widetext}
\begin{equation}\label{eqn:fullintegratedEuler}
R_{s,\text{NL}}(f, \sigma_p)= \frac{1}{6a^2} \left(\frac{4 f^2 a^5+45f^2a -90fa^3 +45a^5 - 15(a^2+3)(f-a^2)^2\arctan(a)}{(f-a)^2\arctan(a) +\frac{1}{3}ab(f(a^2-3)+6a^2)} \right).
\end{equation}
\end{widetext}
The latter was integrated using Mathematica. 

\section{Gaussian FoG effects \label{appn:gaussianfog}}

In this Appendix, for completeness we examine the effects of using a Gaussian function to describe the Finger-of-God smoothing, in place of the Lorentzian used in the main text (Equation \ref{eqn:fingerofgod}). As already remarked in the main text, $\sigma_p$ is not a real velocity dispersion, however, the Finger-of-God smoothing represents a prescription to encapsulate non-linearities.
Using the Gaussian form modifies the power spectrum equation to:

\begin{equation}\label{eqn:fingerofGodgaussian}
P_{s,\text{G}}(k, \mu)= (1+f \mu^2)^2\frac{ P_{\text{ini}}(k)}{e^{(k^2 \mu^2\sigma_p^2/2)}}.
\end{equation}

In this case, the analogous expression to Eq. \ref{eqn:fullintegratedEuler} for the integrated ratio comes to (again, with $a= {k \sigma_p}/{\sqrt{2}}$):

\begin{widetext}
\begin{equation}\label{eqn:fullintegratedGauss}
\begin{aligned}
R_{s,\text{G}}(f, \sigma_p)= \frac{-1}{a^2} \left(\frac{\sqrt{\pi}\left(8a^6+8a^4f-12a^4+6a^2f^2-36a^2f-45f^2\right)\text{erf}(a)}{\sqrt{\pi}\left(4a^4+4a^2f+3f^2\right)\text{erf}(a) - 2a e^{-a^2} f(2a^2(f+2)+3f)} \right) + \\
\frac{-1}{a^2} \left(\frac{e^{-a^2}\left(8a^5(2f^2+4f+3)+48a^3f^2+72a^3f+90af^2+2160a^4f^3\right)}{\sqrt{\pi}\left(4a^4+4a^2f+3f^2\right)\text{erf}(a) - 2a e^{-a^2} f(2a^2(f+2)+3f)} \right).
\end{aligned}
\end{equation}
\end{widetext}

\begin{figure}[htp]
\centering
  \includegraphics[width=0.9\linewidth]{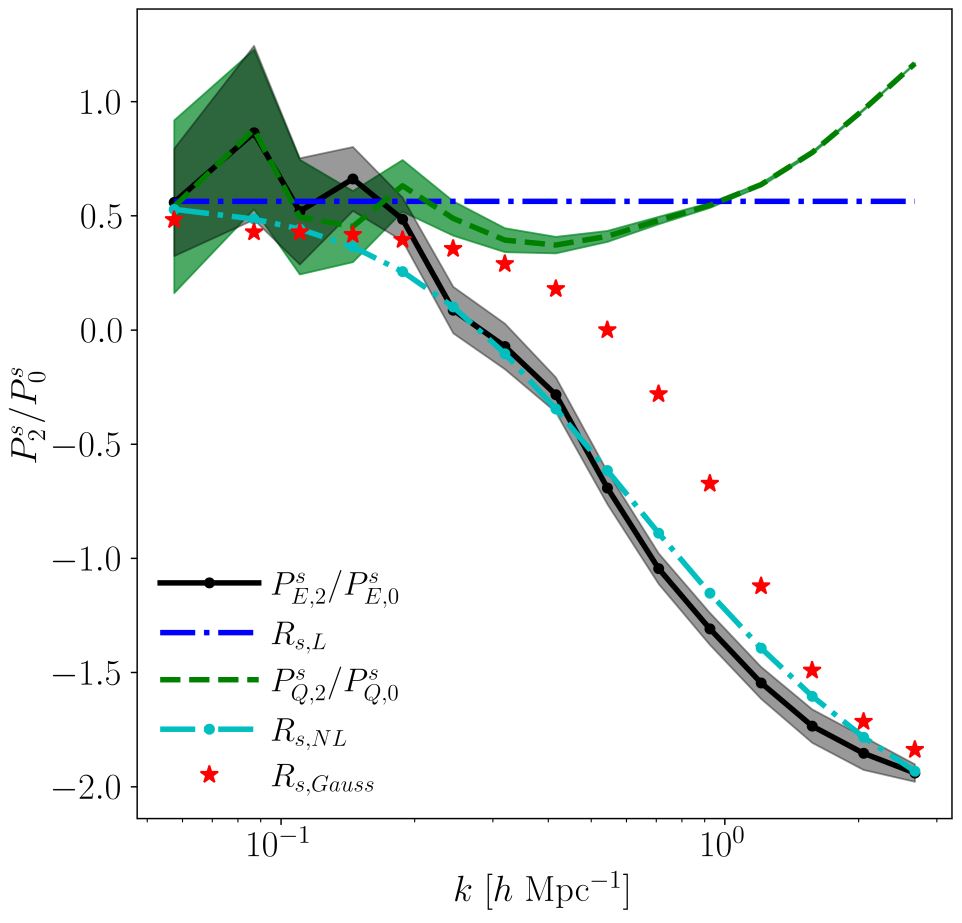} \includegraphics[width=0.9\linewidth]{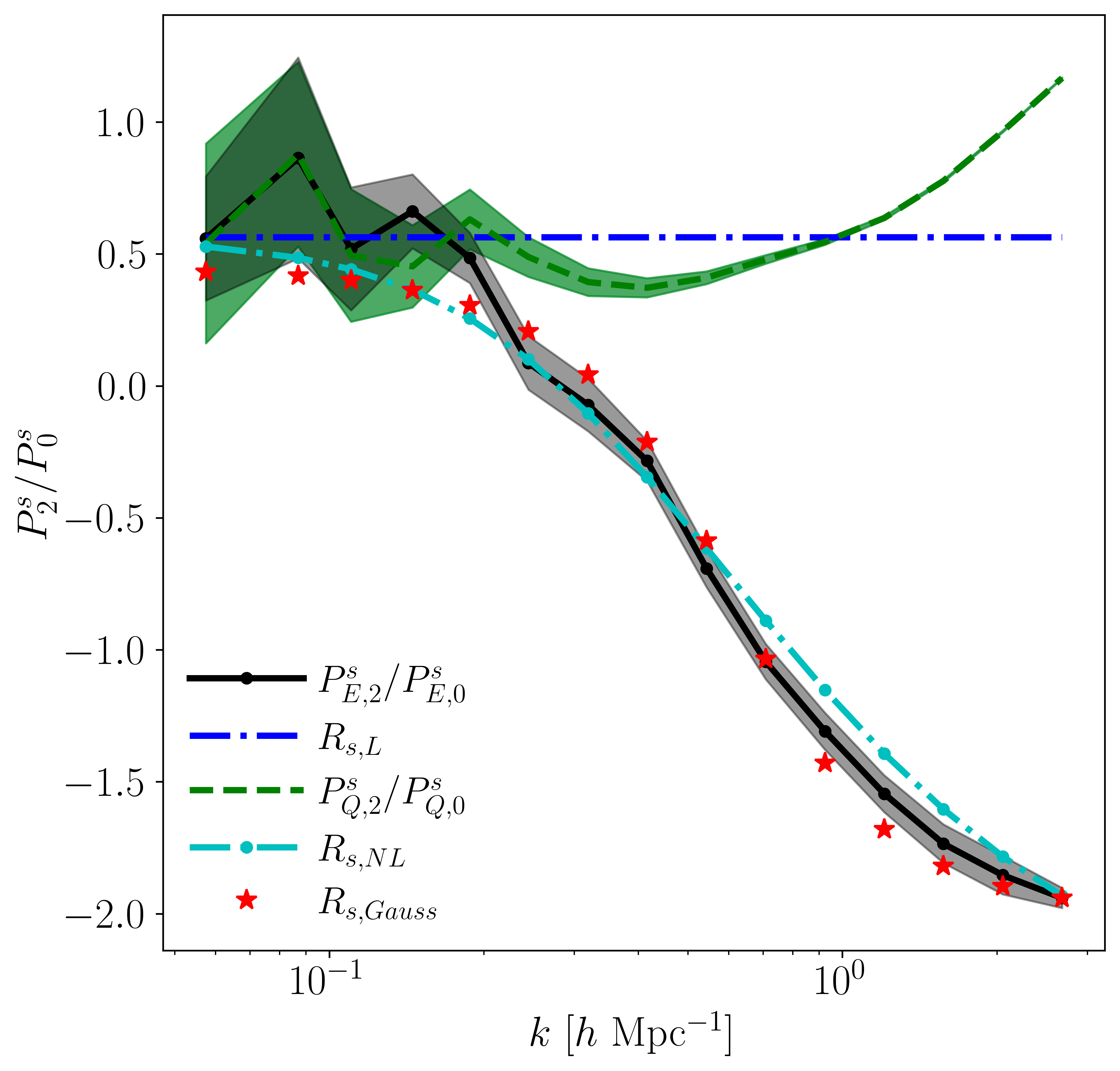}
  \caption{{\it Top panel}: Same as Fig. \ref{fig:EulerianFit}, with the Gaussian Finger-of-God calculation (with the best-fitting $\sigma_p = 373.8$ km/s found in  the main text) shown as red stars. {\it Lower panel}: Same as top panel, but now with the red stars denoting the Gaussian analysis with the re-fitted value of $\sigma_p$ (= 254.6 km/s).} 
  \label{fig:EulerianFitGaussian}
\end{figure}

The ratio of the quadrupole to monopole moment in the Gaussian case, using the best-fitting value of $\sigma_p$  ( = 373.8 km/s) derived in the main text, is shown in the top panel of Fig. \ref{fig:EulerianFitGaussian}. We find a prolonged period of linearity in the Gaussian fit, which is closer to the Lagrangian power spectral outcome (green dashed line) for mid-high scales. 

We also perform a re-fitting for $\sigma_p$ in the Gaussian case, finding it to be $254.6$ km/s. This is different from the corresponding value for the Lorentzian fit, and we note that the lower value of $\sigma_p$ is likely a consequence of the steeper fall-off of the Gaussian function. The corresponding ratio of quadrupole-to-monopole moment is shown in the lower panel of Fig. \ref{fig:EulerianFitGaussian}. The goodness-of-fit, as measured by the difference of normalized squared differences, is found to be marginally better for the Lorentzian fit compared to the Gaussian one (by about 5\%).

\end{document}